\documentclass{ifacconf}

\usepackage{graphicx}      
\usepackage{natbib}        
\usepackage{amsmath}
\usepackage{amssymb}
\usepackage{stackengine}
\usepackage{url}

\DeclareMathOperator*{\argmin}{arg\;min}

\begin{document}
\begin{frontmatter}

\title{On the Initialization of Nonlinear LFR Model Identification with the Best Linear Approximation} 

\thanks[footnoteinfo]{Maarten Schoukens is supported by the European Union's Horizon 2020 research and innovation programme under the Marie Sklodowska-Curie Fellowship (grant agreement nr. 798627). R. T\'oth is supported by the European Research Council (ERC) under the European Union's Horizon 2020 research and innovation programme (grant agreement nr. 714663).}

\author[First]{Maarten Schoukens} 
\author[First]{Roland T\'oth}

\address[First]{Control Systems Group, Dept. of Electrical Eng., Eindhoven University of Technology, Postbus 513, 5600 MB Eindhoven, The Netherlands.(e-mail: \{m.schoukens,r.toth\}@tue.nl).}

\begin{abstract}                
    Balancing the model complexity and the representation capability towards the process to be captured remains one of the main challenges in nonlinear system identification. One possibility to reduce model complexity is to impose structure on the model representation. To this end, this work considers the linear fractional representation framework. In a linear fractional representation the linear dynamics and the system nonlinearities are modeled by two separate blocks that are interconnected with one another. This results in a structured, yet flexible model structure. Estimating such a model directly from input-output data is not a trivial task as the involved optimization is nonlinear in nature. This paper proposes an initialization scheme for the model parameters based on the best linear approximation of the system and shows that this approach results in high quality models on a set of benchmark data sets.
\end{abstract}


\end{frontmatter}

\section{Introduction}
    While the real world is nonlinear and time-varying, so far, we often treated it as linear and time invariant (LTI) when we model and control real-life systems. However, due to, for instance, increasing performance demands, introduction of more light-weight structures and increasing constraints on energy consumption of systems, nonlinear models and nonlinear control has become increasingly important \citep{Schoukens2019}.
    
    A wide range of nonlinear modelling approaches and frameworks is available to the user: a good overview of the variety of available methods is given by \citep{Schoukens2019}. Including structure and prior knowledge has proven to be of key importance to obtain high-quality nonlinear models. While linear models are represented by a hyperplane in a high-dimensional regression space, nonlinear models are a manifold in this high-dimensional space. Including structure in the considered model class reduces the complexity of the modelling problem and makes it a tractable task.
    
    Structure can be imposed or included in the considered model class in various ways. It can be included as a prior in kernel based regression \citep{Pillonetto2011,Birpoutsoukis2017}. Alternatively, the nonlinear dynamics can be represented as an interconnection of LTI blocks and static nonlinear blocks as is done in nonlinear block-oriented modelling approaches \citep{Giri2010,SchoukensM2017b}. Another possibility is to mine a nonlinear input-output relation out of the data, while trading of model complexity with model accuracy \citep{Khandelwal2019}. Structure can also be imposed as a second step in the identification algorithm as in \citep{Esfahani2018} where first a fully coupled nonlinear state-space model is estimated and the structure is only imposed in a second step by restricting the input dimension of the nonlinear state and output mapping using tensor decomposition techniques.
    
\begin{figure}[bt]
		\centering
			\includegraphics[width=0.75\columnwidth]{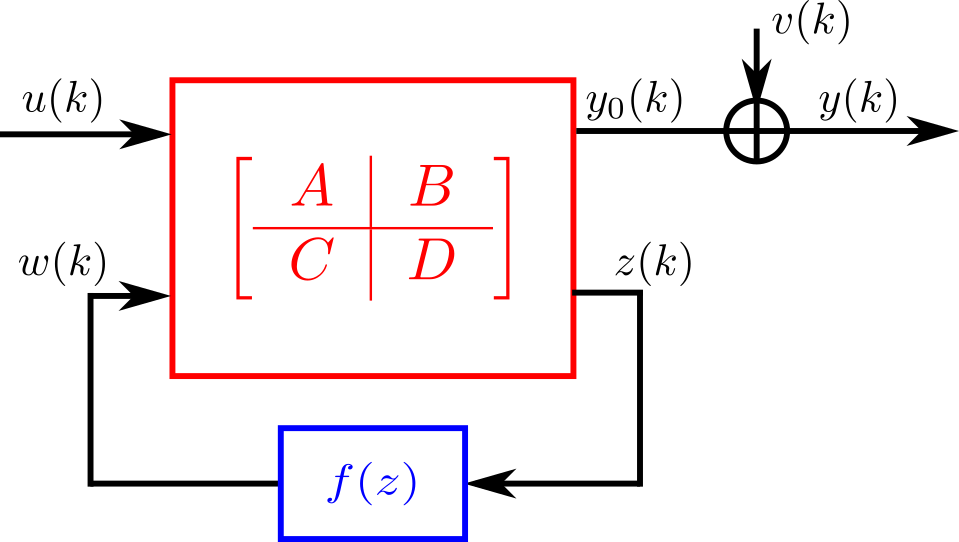}
		\caption{The considered NL-LFR structure represented by MIMO LTI block interconnected with the MIMO static nonlinear function $f(z)$.}
		\label{fig:LFR}
	\end{figure}

    This paper considers models that are composed by the interconnection of a linear fractional representation (LFR) with a static nonlinearity, resulting in the NL-LFR model class (see Figure~
    \ref{fig:LFR}). This model class can be seen as a very general block-oriented structure. Just like block-oriented structures, the nonlinear dynamics are represented as an interconnection between an LTI block and a static nonlinear block. However the inner dynamics of the multiple-input multiple-output (MIMO) LTI block are not imposed. This allows for more flexibility compared to the more common Hammerstein and Wiener block-oriented structures \citep{SchoukensM2017b}. The NL-LFR model class also has a direct link with robust control and linear parameter-varying control design approaches \citep{Zhou1996,Toth2010,SchoukensM2018}.
    
    The identification of NL-LFR structures has already been considered in previous publications. Some approaches derive the NL-LFR model starting from other model structures such as a Volterra series model \citep{Vandersteen1999}, or a nonlinear state-space model \citep{Mulders2013}. Other approaches assume the knowledge of the LTI part of the NL-LFR model and focus on the identification of the static nonlinearity of the model \citep{Hsu2008,Novara2011}. Finally \citep{Vanbeylen2013} starts the identification of the NL-LFR model from multiple linear approximations of the nonlinear systems. The main reason for the various restrictions of the considered identification problem, or on the required prior information (Volterra model, state-space model, multiple linear approximations) is due to the complexity of the involved parameter estimation problem.
    
    This paper proposes an initialization approach of the nonlinear optimization scheme for the estimation of the NL-LFR structure that only requires a single linear approximation of the nonlinear system, which can be obtained using the best linear approximation (BLA) framework \citep{Pintelon2012}. This relaxes the assumed prior knowledge and data requirements of the identification algorithm significantly compared to the prior work available in the literature. The validity of the proposed approach is illustrated on two benchmark examples: the Bouc-Wen hysteretic system benchmark \citep{Noel2016} and the parallel Wiener-Hammerstein benchmark system \citep{SchoukensM2015a}.
    
    The remainder of this paper first discusses the considered NL-LFR model structure in Section~\ref{sec:LFR}. The BLA is introduced in Section~\ref{sec:BLA}. Next, the proposed identification approach is discussed in Section~\ref{sec:Ident}. Finally, the results on the considered benchmark systems are discussed in Section~\ref{sec:results}.

\section{Nonlinear LFR Model Structure} \label{sec:LFR}
	
    \subsection{Model Structure}
        The considered discrete-time nonlinear LFR model structure consists of a, possibly multiple input multiple output (MIMO), static nonlinearity interconnected with the linear fractional representation, as is shown in Figure~\ref{fig:LFR}. In this paper, the linear dynamics are represented using a state-space representation describing the dynamic relation between the LFR inputs $u(k) \in \mathbb{R}^{n_u \times 1}$, $w(k) \in \mathbb{R}^{n_w \times 1}$ and the outputs $y(k) \in \mathbb{R}^{n_y \times 1}$, $z(k) \in \mathbb{R}^{n_z \times 1}$, where $k$ denotes the sample index. The static nonlinearity is represented by a feedforward neural network with one hidden layer with $n_n$ neurons using a nonlinear activation function $\sigma(\cdot)$ and a linear output layer. This results in the following model equations:
        \begin{align} \begin{split} \label{eq:LFR-SS}
            x(k+1) &= A x(k) + \begin{bmatrix}B_u & B_w\end{bmatrix} \begin{bmatrix}u(k) \\ w(k)\end{bmatrix} \\
            \begin{bmatrix}z(k) \\ y_0(k)\end{bmatrix} &= \begin{bmatrix}C_z \\ C_y\end{bmatrix}x(k) + \begin{bmatrix} D_{zu} & 0 \\ D_{yu} & D_{yw} \end{bmatrix} \begin{bmatrix}u(k) \\ w(k)\end{bmatrix}
        \end{split} \end{align}
        and
        \begin{align} \label{eq:LFR-NL}
            w(k) = W_w \sigma \left(W_z z(k) + b_z \right) + b_w,
        \end{align}
        where $\sigma(\cdot)$ is the nonlinear activation function which is often chosen as the hyperbolic tangent function or the radial basis function and $W_z \in \mathbb{R}^{n_n \times n_z}$ and $W_w \in \mathbb{R}^{n_w \times n_n}$ are the inner and outer weights respectively and $b_z \in \mathbb{R}^{n_n \times 1}$ and $b_w \in \mathbb{R}^{n_w \times 1}$ are the respective biases of the neural network. The states are represented by $x(k) \in \mathbb{R}^{n_x \times 1}$, while the matrices $A  \in \mathbb{R}^{n_x \times n_x}$, $B_u \in \mathbb{R}^{n_x \times n_u}$, $B_w \in \mathbb{R}^{n_x \times n_w}$, $C_y \in \mathbb{R}^{n_y \times n_x}$, $C_z \in \mathbb{R}^{n_z \times n_x}$, $D_{zu} \in \mathbb{R}^{n_z \times n_u}$, $D_{yu} \in \mathbb{R}^{n_y \times n_u}$ and $D_{yw} \in \mathbb{R}^{n_y \times n_w}$ correspond to the parameters to be estimated together with the weights in eq.~\eqref{eq:LFR-NL}. Note that the term $D_{zw}$ is not present in the considered state-space equation. This prevents the presence of algebraic equations in the model expression and hence avoiding well-posedness problems. 

        The nonlinear modelling capabilities of the NL-LFR structure can be tuned by increasing or decreasing the dimension of the $z$ and $w$ signals. This allows the model structure to go from very structured (only one SISO static nonlinearity present in the model) to rather unstructured (a high dimensional MIMO static nonlinearity). Observe as well that many of the commonly used block-oriented model structures such as the Wiener, Hammerstein and Hammerstein-Wiener model structures are subsets of the NL-LFR model structure. Furthermore by making use of a state-space representation of the dynamics, the model scales easily towards multiple inputs $u$ and outputs $y$.
	
	    A zero-mean, possibly colored, additive noise source $v(k)$ is assumed to present at the output $y(k)$ only. The noise source is assumed to have a finite variance. In case the noise is white, this corresponds to the classical output-error noise framework.
	
	\subsection{Uniqueness of the Parametrization}
	    The considered model representation is not unique. Beyond the well-known arbitrary state-space transformation that defines an equivalence class of models around \eqref{eq:LFR-SS}-\eqref{eq:LFR-NL}, also the neural network representation of the static nonlinearity is not uniquely parametrized. A simple permutation of the inner and other weights and biases can result in the same input-output behavior of the neural network \eqref{eq:LFR-NL}. Beyond the non-uniqueness of the LTI and static nonlinear blocks separately, a linear gain can also be exchanged between the static nonlinear and the LTI block \citep{SchoukensM2017b}.

\section{Best Linear Approximation} \label{sec:BLA}
    The Best Linear Approximation (BLA) of a nonlinear system is an LTI approximation of the input-output map of the system, in a mean square sense \citep{Pintelon2012, Enqvist2005}. The BLA is obtained as:
    \begin{align} \begin{split} \label{eq:BLA}
        G_{\mathrm{bla}}(q) 	&= \underset{G(q)}{\argmin} \: E_{u,v}\left\{ \left| \tilde{y}(k) - G(q)\tilde{u}(k) \right|^{2} \right\},  \\
				\tilde{u}(k)	&= u(k) - E_{u}\left\{ u(k) \right\}, \\
		 	  \tilde{y}(k)	&= y(k) - E_{u,v}\left\{ y(k) \right\}, 
    \end{split} \end{align}
    where $E_{u,v}\left\{\cdot\right\}$ denotes the expected value operator taken w.r.t. the random variations due to the input $u(k)$ and the output noise $v(k)$, $q^{-1}$ denotes the backwards shift operator. As can be observed in eq.~\eqref{eq:BLA}, the BLA of a nonlinear system is dependent on the properties of the considered input class \citep{Schoukens2015}.
    
\section{Identification of a Nonlinear LFR Model} \label{sec:Ident}
    \subsection{Parameter Estimation}
        The model parameters are obtained as the minimization of the mean squared simulation error:
        \begin{align} \label{eq:cost}
            V_N(\theta) = \frac{1}{N} \sum_{k=1}^N \left( y(k) - \hat{y}(k|\theta) \right)^2,
        \end{align}
        \begin{align} \label{eq:theta}
            \hat{\theta} = \underset{\theta}{\argmin} V_N(\theta),
        \end{align}
        where $\hat{y}(k|\theta)$ is the simulated output of the NL-LFR model given the parameter vector $\theta$. The parameter vector $\theta$ contains all the model parameters: the state-space matrix entries and the weights and biases of the neural network representing the static nonlinearity. $N$ represents the total number of samples over which the cost function is computed.
        
        Since eq.~\eqref{eq:cost} is typically nonlinear in the parameters and not convex, but its gradients can be efficiently computed, a gradient descent-type algorighm, like the Levenberg-Marquardt algorithm \citep{Levenberg1944} is used to minimize the cost function. A 'data-driven coordinate frame' is used to get rid off equivalent gradient directions (resulting in a rank deficient Jacobian) due to the non-uniqueness of the model representation. In practice this is achieved using the singular value decomposition of the Jacobian \citep{Wills2008}.
        
    \subsection{Parameter Initialization}
        The Levenberg-Marquardt algorithm is not guaranteed to converge to the global minimum of the cost function, it converges to the 'closest' local minimum. The use of 'efficient' initial estimates to start the nonlinear optimization can help to guide the optimization algorithm to the global minimum. 
        
        The BLA estimate of a nonlinear system has proven to be a good initial point to start the identification of a nonlinear model \citep{Paduart2010,Vanbeylen2013,SchoukensM2017b}. While the previous BLA-based approach \citep{Vanbeylen2013} requires 2 BLA estimates at 2 different setpoints of the system, this work proposes an initialization procedure starting from only one state-space BLA estimate ($A_{\mathrm{bla}}$, $B_{\mathrm{bla}}$, $C_{\mathrm{bla}}$, $D_{\mathrm{bla}}$).
        
        First the BLA state-space matrices are transformed such that each of the states has a unit variance:
        \begin{align} \begin{split}
            \bar{A}_{\mathrm{bla}} = TA_{\mathrm{bla}}T^{-1}, \quad &\bar{B}_{\mathrm{bla}} = TB_{\mathrm{bla}} \\
            \bar{C}_{\mathrm{bla}} = C_{\mathrm{bla}}T^{-1}, \quad &\bar{D}_{\mathrm{bla}} = D_{\mathrm{bla}} \\
        \end{split} \end{align}
        where $T \in \mathcal{R}^{n_x \times n_x}$ is a diagonal matrix with the inverse of the standard deviation (taken over the samples) of $x$ on its diagonal.
        
        Next the BLA estimates are embedded into the NL-LFR model, while the other parameters are initialized as zero or as a random variable:
        \begin{equation}
            \begin{alignedat}{2}
            &A = \bar{A}_{\mathrm{bla}}  \quad       &&B_{u} = \bar{B}_{\mathrm{bla}}\\
            &C_{y} = \bar{C}_{\mathrm{bla}} \quad    &&D_{yu} = \bar{D}_{\mathrm{bla}} \\
            &B_{w} = 0       \quad                   &&D_{yw}= 0  \\
            &W_{w}\sim \mathcal{U}(-1,1) \quad       &&W_{z} \sim \mathcal{U}(-1,1) \\
            &b_{w}= 0      \quad                     &&b_{z} \sim \mathcal{U}(-1,1) 
            \end{alignedat} 
        \end{equation}
        where $\mathcal{U}(a,b)$ denotes a uniformly distribution with a support from $a$ to $b$. The uniform random initialization of the parameters is the common approach when training a neural network  \citep{Bishop1995}. The initialization of $C_{z}$ and $D_{zu}$ is slightly different:
            \begin{equation}
                \begin{alignedat}{2}
                &C_{*} \sim \mathcal{U}(-1,1) \quad && \\
                &D_{+} \sim \mathcal{U}(-1,1) \quad && D_{*} =  D_{+} T_{u}^{-1}\\
                &z_{*} = C_{*} x + D_{*}u && \\
                &C_{z} = T_{z_{*}}^{-1} C_{*} \quad &&D_{zu} = T_{z_{*}}^{-1} D_{*}
                \end{alignedat} 
            \end{equation}
         where $T_{z_{*}} \in \mathcal{R}^{n_z \times n_z}$ and $T_{u} \in \mathcal{R}^{n_u \times n_u}$ are a diagonal matrices with the inverse of the standard deviation (taken over the samples) of $z_{*}$ and $u$ respectively on its diagonal.     
         
         The transformations $T$, $T_{u}$, $T_{z_{*}}$ ensure that the initial estimates of the $x$ and $z$ signals, which act as the input of the neural network, have a standard deviation equal to one and are zero-mean. This is generally recognized in the neural network literature to improve the estimation of the model parameters \citep{Bishop1995}. The initial model has the same performance as the BLA estimate since $B_{w}$ and $D_{yw}$ are initialized as zero matrices. It also ensures that, if the BLA estimate is stable, the initial estimate of the NL-LFR model is stable as well. 
    
    \subsection{BLA Identification}
        Many different approaches are available in the literature to estimate the BLA. For example, nonparametric and parametric frequency-domain methods \citep{Pintelon2012,Paduart2010} or the prediction error method \citep{Ljung1999} have been commonly used.
    
       This paper directly estimates a state-space model of the BLA using the time-domain prediction error method as implemented in Matlab by the function {\tt ssest} with the signals $u(k)$ and $y(k)$ as the input and output data respectively. This function initializes the parameter estimates using either a subspace approach or an iterative rational function estimation approach. The state-space matrices are refined subsequently using the prediction error minimization approach \citep{Ljung1999}.
       
       When the system is strongly nonlinear, obtaining a sufficiently good estimate of the BLA might not be trivial. The lack of a good BLA estimate can result in sub-optimal NL-LFR model estimates. Worse, for specific setpoints of the system the BLA can be equal to zero \citep{SchoukensM2017b}. The case when the BLA only captures part of the system dynamics will again result in sub-optimal NL-LFR model estimates.

\section{Benchmark Results} \label{sec:results}
    Two benchmark datasets are considered: the Bouc-Wen Hysteretic system \citep{Noel2016} and the parallel Wiener-Hammerstein datasets \citep{SchoukensM2015a}. 
    
    The Bouc-Wen system is a hysteretic system featuring a dynamic nonlinearity: it is a mass-spring-damper system with a hysteretic restoring force. This hysteretic nonlinearity is governed by a differential equation containing hard nonlinearities such as the absolute value operator.
    
    The parallel Wiener-Hammerstein system is obtained as the parallel cascade of two Wiener-Hammerstein systems (a linear input filter followed by a static nonlinearity, followed by a linear output filter). The system has 12th order dynamics ($n_x = 12$), which is challenging for many black-box identification algorithms. The static nonlinearities are realized by a diode-resistor network resulting in a one-sided and a two-sided saturation nonlinearity.
    
    \subsection{Bouc-Wen Hysteretic System}
        \subsubsection{The Data:} 
            The Bouc-Wen system is available as a simulation script. This allows the users to generate their own data for identification. We used one period of a random phase multisine input signal \citep{Pintelon2012}. The signal was $N=8192$ samples long and it excites the full frequency grid between 5 Hz and 150 Hz. The input signal amplitude is 50 N\textsubscript{rms}.
            
            Two test datasets are available for the Bouc-Wen benchmark: a multisine and a sinesweep dataset. The multisine test output is obtained as the steady-state response of the system excited by a random phase multisine of $N=8192$ samples long, exciting the full frequency grid between 5 and 150 Hz, with a signal amplitude of 50 N\textsubscript{rms}. The sinesweep test output is obtained by exciting the system, starting from zero initial conditions, with a sinesweep signal with an amplitude of 40  N\textsubscript{rms}. The frequency band from 20 to 50 Hz is covered at a sweep rate of 10 Hz/min.
        
        \subsubsection{Model Settings:}
            In line with previous finding the linear dynamics are described by a 3rd order state-space model ($n_x = 3$) \citep{Noel2017b}. From the system description \citep{Noel2016} it can be concluded that the 2 $z$-variables and one $w$-variable should result in a suitable model structure to capture the system dynamics. However, for illustrative purposes, three cases are considered here: \{$n_z = 1$, $n_w = 1$\}, \{$n_z = 2$, $n_w = 1$\} and \{$n_z = 2$, $n_w = 2$\}. A total of 15 neurons ($n_n = 15$, tansig activation function) are used to represent the static nonlinearity.
        
        \subsubsection{Results:}
            
            As a measure of model quality, the simulation RMSE (root mean squared error) on the multisine and sinesweep test dataset is reported:
            \begin{align} \label{eq:RMSE}
                e_{\text{RMSE}} = \sqrt{\frac{1}{N} \sum_{k=1}^N \left( y(k) - \hat{y}(k|\theta) \right)^2},
            \end{align}
            where $y(k)$ is the observed test output and $\hat{y}(k|\theta)$ is the simulated output using the estimated model. To make sure the model output is in steady state for the multisine dataset two periods are simulated and the RMSE is calculated on the second period. The sinesweep data is not in steady state, hence, the first 2000 samples are ignored when calculating the RMSE to allow the transient to decay.
            
            Table~\ref{table:BW} shows a comparison of the results obtained using an LTI model (the model used to initialize the NL-LFR parameter optimization) and the various NL-LFR cases considered. It can be observed that the most simple NL-LFR configuration is not sufficiently rich to capture the complete nonlinear behaviour of the system, only a factor 3 reduction of the model error is obtained compared to the LTI case. The two cases where $n_z = 2$ on the other hand succeed in reducing the model error with a factor 20 compared to the BLA model, indicating that this is a much more suited model structure for the Bouc-Wen benchmark system. This is indeed in line with the expectations as this matches with the model equations reported in \citep{Noel2016}. Both NL-LFR models with $n_z = 2$ perform similar, which is an indication that the model structure with $n_z = 2$ and $n_w = 1$ is matching best with the underlying system structure.
            
\begin{figure}[bt] 
        		\centering
        			\includegraphics[width=0.9\columnwidth]{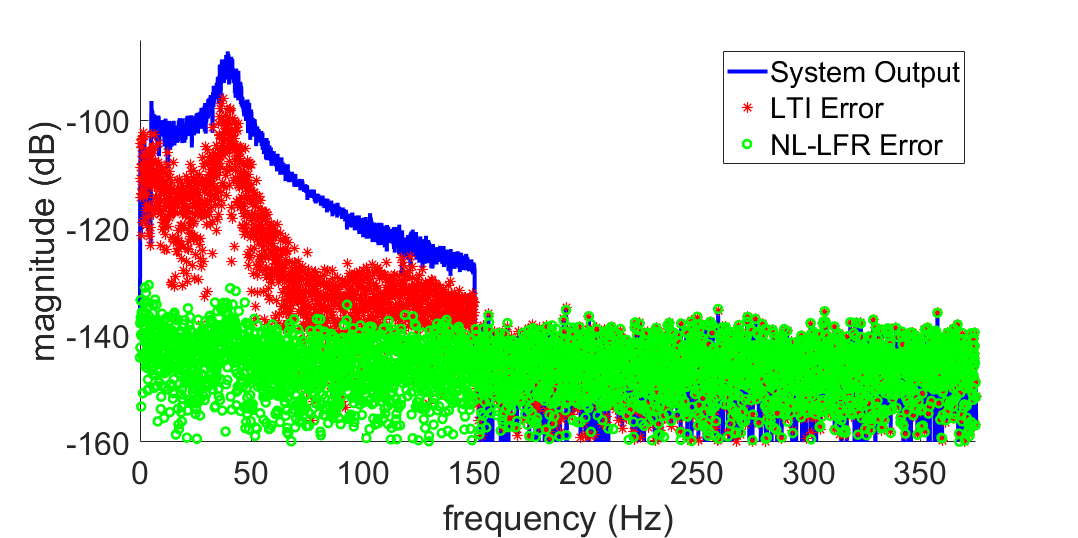}
        		\caption{Frequency-domain validation of the NL-LFR model with $n_z = 2$ and $n_w = 1$ using the Bouc-Wen multisine test data. The true system output is shown in blue, the residuals obtained with a 3rd order LTI model are shown in red and the residuals obtained with the NL-LFR model are shown in green.}
        		\label{fig:BWFreq}
        	\end{figure}

\begin{figure}[bt] 
        		\centering
        			\includegraphics[width=0.9\columnwidth]{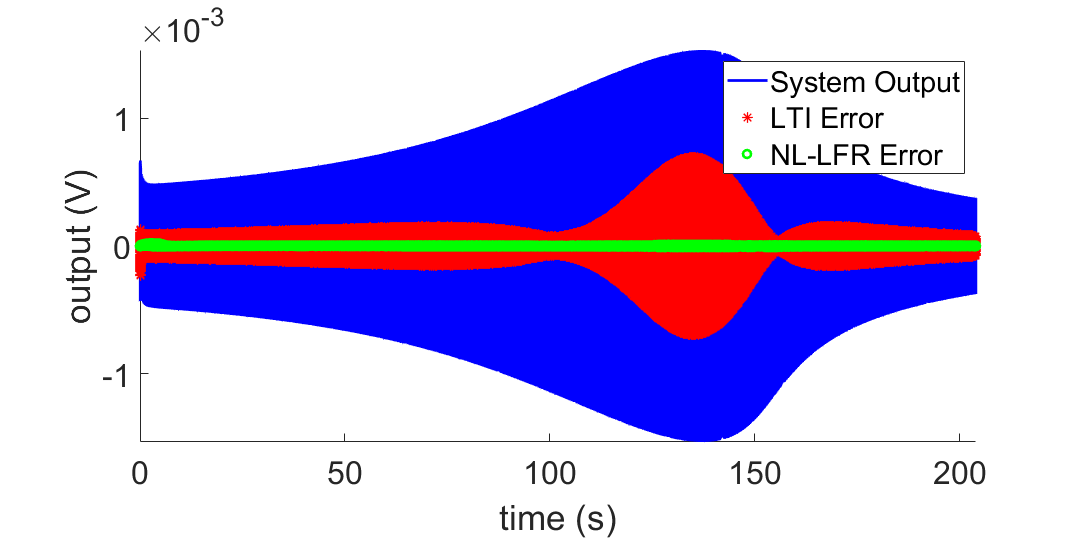}
        		\caption{Time-domain validation of the NL-LFR model with $n_z = 2$ and $n_w = 1$ using the Bouc-Wen sinesweep test data. The true system output is shown in blue, the residuals obtained with an LTI model are shown in red and the residuals obtained with the NL-LFR model are shown in green.}
        		\label{fig:BWTime}
        	\end{figure}

            The test results for the LTI model and the NL-LFR model with $n_z = 2$ and $n_w = 1$ are shown in Figures~\ref{fig:BWFreq} and \ref{fig:BWTime}. It is apparent from the figures that the NL-LFR model significantly outperforms the LTI model. The signal and error behavior outside the excited frequency range indicate that the obtained model error is very close to the noise floor. The quality of the results are, to the authors knowledge among the best black-box identification results obtained on this dataset so far, the RMSE is 2-3 times lower than the one reported in \citep{Esfahani2018}.
            
            \addtolength{\tabcolsep}{-1pt} 
            \begin{table}[ht]
                \centering
                \caption{Bouc-Wen benchmark results: simulation RMSE. The first column shows the results obtained with a 3rd order LTI model, the next three columns show the results obtained with the NL-LFR for different $n_z$ and $n_w$ values. The final column shows the results that are reported in \citep{Esfahani2018} using a decoupled polynomial nonlinear state-space approach. The rms errors \eqref{eq:RMSE} are reported in m\textsubscript{rms}.}
                \begin{tabular}{c | c c c c c } 
                 \rule{0pt}{2ex} 
                           &     & $n_z = 1$ & $n_z = 2$  & $n_z = 2$  & decoupled\\ 
                           & LTI  & $n_w = 1$ & $n_w = 1$  & $n_w = 2$ & PNLSS\\[0.5ex] 
                 \hline 
                 Multisine & $15.8e^{-5}$ & $5.31e^{-5}$ & $0.72e^{-5}$ & $0.74e^{-5}$ & $1.34e^{-5}$ \\ 
                 Sinesweep & $17.7e^{-5}$ & $4.19e^{-5}$ & $0.32^{-5}$  & $0.56e^{-5}$ & $1.12e^{-5}$
                \end{tabular}
                \label{table:BW}
            \end{table}
            \addtolength{\tabcolsep}{1pt} 
            
    \subsection{Parallel Wiener-Hammerstein System}
        \subsubsection{The Data:}
            A detailed discussion of the estimation and test data for the parallel Wiener-Hammerstein system is given in \citep{SchoukensM2015a}. This work used a subset of the available estimation data: 1 steady-state period of a random phase multisine signal of 16384 samples long at 5 different input amplitudes. The test data consists of another multisine realization at each of the 5 different amplitudes and a random Gaussian noise sequence with a linearly growing amplitude over time (denoted as the arrow signal later on).
        
        \subsubsection{Model Settings:}
            As described in \citep{SchoukensM2015a} the linear dynamics are described by a 12th order state-space model ($n_x = 12$) and 2 parallel nonlinear branches are present in the system. This indicates 2 $z$-variables and 2 $w$-variable should result in the 'correct' model structure. A total of 15 neurons ($n_n = 15$, tansig activation functions) are used to represent the static nonlinearity.
            
        \subsubsection{Results:}  
            Table~\ref{table:pWH} shows a comparison of the results obtained using an LTI model and the obtained NL-LFR model for the 6 different test signals. Also the results obtained using the parallel Wiener-Hammerstein method described in \citep{SchoukensM2015a} are shown for comparison. Similar to the previous case study, to make sure the model output for both the multisine and arrow dataset is in steady state two periods are simulated, the RMSEs are calculated on the second period. It can be observed that the NL-LFR model outperforms the LTI models even though the LTI model has been specifically trained for each of the multisine input amplitudes separately.
            
\begin{figure}[bt] 
        		\centering
        			\includegraphics[width=0.9\columnwidth]{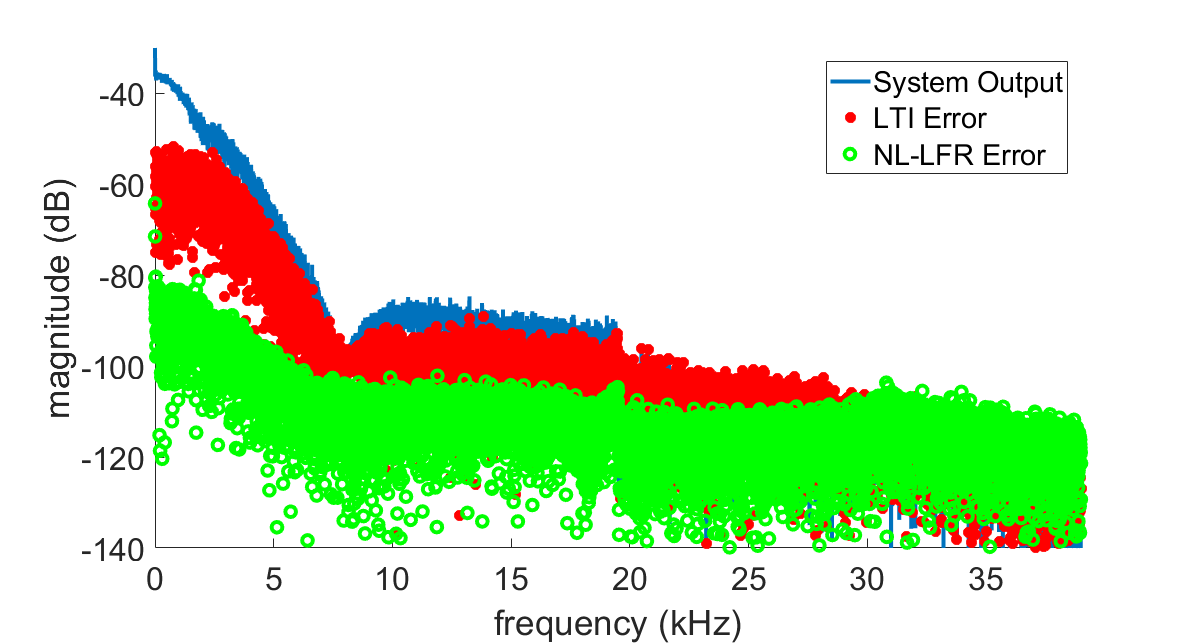}
        		\caption{Frequency-domain validation of the NL-LFR model using the parallel Wiener-Hammerstein multisine test data. The true system output is shown in blue, the residuals obtained with a 12th order LTI model are shown in red and the residuals obtained with the NL-LFR model are shown in green.}
        		\label{fig:pWHFreq}
        	\end{figure}

\begin{figure}[bt] 
        		\centering
        			\includegraphics[width=0.9\columnwidth]{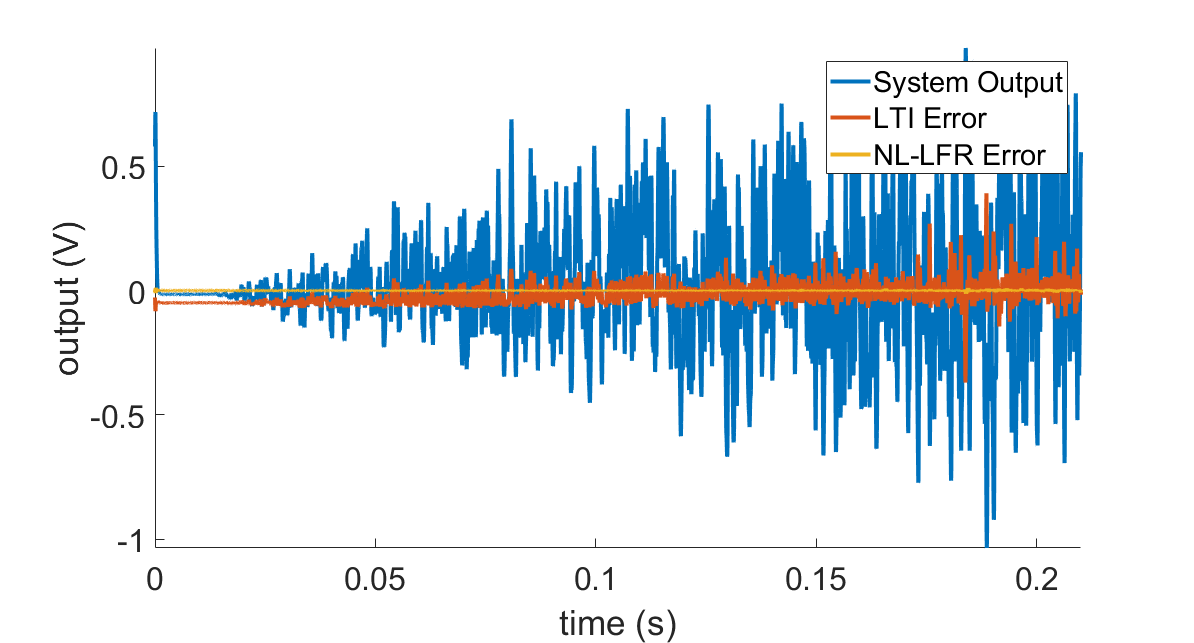}
        		\caption{Time-domain validation of the NL-LFR model using the parallel Wiener-Hammerstein arrow test data. The true system output is shown in blue, the residuals obtained with a 12th order LTI model are shown in red and the residuals obtained with the NL-LFR model are shown in green.}
        		\label{fig:pWHTime}
        	\end{figure}

            The test results for the LTI model and the NL-LFR model on the largest multisine amplitude and on the arrow signal are shown in Figures~\ref{fig:pWHFreq} and \ref{fig:pWHTime}. It is apparent from the figures that the NL-LFR model significantly outperforms the LTI model.
            
            The obtained results are in line with the results reported in \citep{SchoukensM2015a} where a parallel Wiener-Hammerstein model has been fitted on the data. Even though the specific Wiener-Hammerstein nature of the system has not been imposed on the NL-LFR model, a similar model quality has been obtained.
            
            \begin{table}[ht]
                \centering
                \caption{The parallel Wiener-Hammerstein benchmark results: simulation RMSE. The first column shows the results obtained with a 12th order LTI model specifically estimated for that amplitude. The LTI model obtained for the 1000 mV\textsubscript{rms} input signal also has been used for the arrow test signal. The second and third column shows the results obtained with the NL-LFR model (this paper) and a parallel Wiener-Hammerstein approach \citep{SchoukensM2015a}. The rms errors \eqref{eq:RMSE}  are reported in V\textsubscript{rms}.}
                \begin{tabular}{c | c c c} 
                 \rule{0pt}{2ex} 
                           & LTI  & NL-LFR & pWH \\
                 \hline 
                 Multisine 100 mV\textsubscript{rms}  & $0.83e^{-3}$ & $0.30e^{-3}$ & $0.30e^{-3}$ \\
                 Multisine 325 mV\textsubscript{rms}  & $9.46e^{-3}$ & $0.51e^{-3}$ & $0.50e^{-3}$\\
                 Multisine 550 mV\textsubscript{rms}  & $19.9e^{-3}$ & $0.72e^{-3}$ & $0.38e^{-3}$\\
                 Multisine 775 mV\textsubscript{rms}  & $30.7e^{-3}$ & $1.03e^{-3}$ & $0.57e^{-3}$\\
                 Multisine 1000 mV\textsubscript{rms}  & $37.7e^{-3}$ & $1.37e^{-3}$ & $1.10e^{-3}$\\
                 Arrow Signal & $45.5e^{-3}$ & $1.42e^{-3}$ & $2.66e^{-3}$ 
                \end{tabular}
                 \label{table:pWH}
            \end{table}
            
    \subsection{Discussion}
        The obtained results on the Bouc-Wen and parallel Wiener-Hammerstein benchmark datasets illustrate the versatile nature of the NL-LFR model structure and of the proposed identification algorithm. High-quality modelling results comparable with the state-of-the-art have been obtained even though the two systems that were considered here are significantly different in nature. Note as well that due to the structured nature of the NL-LFR, it has little problem in handling nonlinear systems with high-order dynamics such as the parallel Wiener-Hammerstein system if the dimension of the $z$ and $w$ signals can be kept under control.

\section{Conclusion and Future Work}
    The problem of identifying an NL-LFR model of a nonlinear system has been addressed in this paper. The NL-LFR model structure offers a flexible, yet structured model class which is a generalization of the popular block-oriented Hammerstein and Wiener model class. The proposed initialization of the nonlinear-in-the-parameters non-convex optimization associated with the identification problem, starting from a BLA estimate of the system, has proven to be successful on the two considered benchmark problems. A detailed theoretical analysis of the proposed identification approach is currently lacking and will be the subject of future work.

\bibliography{Bibliography}             

\begin{thebibliography}{26}
\providecommand{\natexlab}[1]{#1}
\providecommand{\url}[1]{\texttt{#1}}
\providecommand{\urlprefix}{URL }
\expandafter\ifx\csname urlstyle\endcsname\relax
  \providecommand{\doi}[1]{doi:\discretionary{}{}{}#1}\else
  \providecommand{\doi}{doi:\discretionary{}{}{}\begingroup
  \urlstyle{rm}\Url}\fi

\bibitem[{Birpoutsoukis et~al.(2017)Birpoutsoukis, Marconato, Lataire, and
  Schoukens}]{Birpoutsoukis2017}
Birpoutsoukis, G., Marconato, A., Lataire, J., and Schoukens, J. (2017).
\newblock Regularized nonparametric volterra kernel estimation.
\newblock \emph{Automatica}, 82, 324 -- 327.

\bibitem[{Bishop(1995)}]{Bishop1995}
Bishop, C.M. (1995).
\newblock \emph{Neural Networks for Pattern Recognition}.
\newblock Oxford University Press, Inc., New York, NY, USA.

\bibitem[{Enqvist(2005)}]{Enqvist2005}
Enqvist, M. (2005).
\newblock \emph{{Linear Models of Nonlinear systems}}.
\newblock Ph.D. thesis, Institute of technology, Link\"{o}ping University,
  Sweden.

\bibitem[{{Fakhrizadeh Esfahani} et~al.(2018){Fakhrizadeh Esfahani}, Dreesen,
  Tiels, {No\"el}, and Schoukens}]{Esfahani2018}
{Fakhrizadeh Esfahani}, A., Dreesen, P., Tiels, K., {No\"el}, J.P., and
  Schoukens, J. (2018).
\newblock Parameter reduction in nonlinear state-space identification of
  hysteresis.
\newblock \emph{Mechanical Systems and Signal Processing}, 104, 884--895.

\bibitem[{Giri and Bai(2010)}]{Giri2010}
Giri, F. and Bai, E. (eds.) (2010).
\newblock \emph{{Block-oriented Nonlinear System Identification}}, volume 404
  of \emph{Lecture Notes in Control and Information Sciences}.
\newblock Springer-Verlag, London.

\bibitem[{Hsu et~al.(2008)Hsu, Poolla, and Vincent}]{Hsu2008}
Hsu, K., Poolla, K., and Vincent, T.L. (2008).
\newblock Identification of structured nonlinear systems.
\newblock \emph{IEEE Transactions on Automatic Control}, 53(11), 2497--2513.

\bibitem[{Khandelwal et~al.(2019)Khandelwal, Schoukens, and
  Toth}]{Khandelwal2019}
Khandelwal, D., Schoukens, M., and Toth, R. (2019).
\newblock Data-driven modelling of dynamical systems using tree adjoining
  grammar and genetic programming.
\newblock In \emph{2019 IEEE Congress on Evolutionary Computation, CEC 2019 -
  Proceedings}, 2673--2680. Wellington, New Zealand.

\bibitem[{Levenberg(1944)}]{Levenberg1944}
Levenberg, K. (1944).
\newblock {A method for the solution of certain problems in least squares}.
\newblock \emph{Quarterly of Applied Mathematics}, 2, 164--168.

\bibitem[{Ljung(1999)}]{Ljung1999}
Ljung, L. (1999).
\newblock \emph{{System Identification: Theory for the User (second edition)}}.
\newblock Prentice Hall, Upper Saddle River, New Jersey.

\bibitem[{No\"el et~al.(2017)No\"el, Esfahani, Kerschen, and
  Schoukens}]{Noel2017b}
No\"el, J.P., Esfahani, A., Kerschen, G., and Schoukens, J. (2017).
\newblock A nonlinear state-space approach to hysteresis identification.
\newblock \emph{Mechanical Systems and Signal Processing}, 84(B), 171--184.

\bibitem[{No\"el and Schoukens(2016)}]{Noel2016}
No\"el, J.P. and Schoukens, M. (2016).
\newblock {Hysteretic benchmark with a dynamic nonlinearity}.
\newblock In \emph{Workshop on Nonlinear System Identification Benchmarks},
  7--14. Brussels, Belgium.

\bibitem[{Novara et~al.(2011)Novara, Vincent, Hsu, Milanese, and
  Poolla}]{Novara2011}
Novara, C., Vincent, T.L., Hsu, K., Milanese, M., and Poolla, K. (2011).
\newblock Parametric identification of structured nonlinear systems.
\newblock \emph{Automatica}, 47, 711--721.

\bibitem[{Paduart et~al.(2010)Paduart, Lauwers, Swevers, Smolders, Schoukens,
  and Pintelon}]{Paduart2010}
Paduart, J., Lauwers, L., Swevers, J., Smolders, K., Schoukens, J., and
  Pintelon, R. (2010).
\newblock {Identification of nonlinear systems using polynomial nonlinear state
  space models}.
\newblock \emph{Automatica}, 46(4), 647--656.

\bibitem[{{Pillonetto} et~al.(2011){Pillonetto}, {Quang}, and
  {Chiuso}}]{Pillonetto2011}
{Pillonetto}, G., {Quang}, M.H., and {Chiuso}, A. (2011).
\newblock A new kernel-based approach for nonlinear system identification.
\newblock \emph{IEEE Transactions on Automatic Control}, 56(12), 2825--2840.

\bibitem[{Pintelon and Schoukens(2012)}]{Pintelon2012}
Pintelon, R. and Schoukens, J. (2012).
\newblock \emph{{System Identification: A Frequency Domain Approach}}.
\newblock Wiley-IEEE Press, Hoboken, New Jersey, 2nd edition.

\bibitem[{{Schoukens} and {Ljung}(2019)}]{Schoukens2019}
{Schoukens}, J. and {Ljung}, L. (2019).
\newblock Nonlinear system identification: A user-oriented road map.
\newblock \emph{IEEE Control Systems Magazine}, 39(6), 28--99.

\bibitem[{Schoukens et~al.(2015{\natexlab{a}})Schoukens, Pintelon, Rolain,
  Schoukens, Tiels, Vanbeylen, {Van Mulders}, and Vandersteen}]{Schoukens2015}
Schoukens, J., Pintelon, R., Rolain, Y., Schoukens, M., Tiels, K., Vanbeylen,
  L., {Van Mulders}, A., and Vandersteen, G. (2015{\natexlab{a}}).
\newblock {Structure discrimination in block-oriented models using linear
  approximations: A theoretic framework}.
\newblock \emph{Automatica}, 53, 225--234.

\bibitem[{Schoukens et~al.(2015{\natexlab{b}})Schoukens, Marconato, Pintelon,
  Vandersteen, and Rolain}]{SchoukensM2015a}
Schoukens, M., Marconato, A., Pintelon, R., Vandersteen, G., and Rolain, Y.
  (2015{\natexlab{b}}).
\newblock {Parametric identification of parallel Wiener-Hammerstein systems}.
\newblock \emph{Automatica}, 51(1), 111--122.

\bibitem[{Schoukens and Tiels(2017)}]{SchoukensM2017b}
Schoukens, M. and Tiels, K. (2017).
\newblock Identification of block-oriented nonlinear systems starting from
  linear approximations: A survey.
\newblock \emph{Automatica}, 85, 272--292.

\bibitem[{Schoukens and T\'oth(2018)}]{SchoukensM2018}
Schoukens, M. and T\'oth, R. (2018).
\newblock From nonlinear identification to linear parameter varying models:
  Benchmark examples.
\newblock In \emph{18th IFAC Symposium on system identification (SYSID)}.
  Stockholm, Sweden.

\bibitem[{T\'{o}th(2010)}]{Toth2010}
T\'{o}th, R. (2010).
\newblock \emph{Modeling and Identification of Linear Parameter-Varying
  Systems}, volume 403 of \emph{Lecture Notes in Control and Information
  Sciences}.
\newblock Springer-Verlag, Berlin Heidelberg.

\bibitem[{{Van Mulders} et~al.(2013){Van Mulders}, Schoukens, and
  Vanbeylen}]{Mulders2013}
{Van Mulders}, A., Schoukens, J., and Vanbeylen, L. (2013).
\newblock {Identification of systems with localised nonlinearity: From
  state-space to block-structured models}.
\newblock \emph{Automatica}, 49(5), 1392--1396.

\bibitem[{Vanbeylen(2013)}]{Vanbeylen2013}
Vanbeylen, L. (2013).
\newblock {Nonlinear LFR Block-Oriented Model: Potential Benefits and Improved,
  User-Friendly Identification Method}.
\newblock \emph{IEEE Transactions on Instrumentation and Measurement}, 62(12),
  3374--3383.

\bibitem[{Vandersteen and Schoukens(1999)}]{Vandersteen1999}
Vandersteen, G. and Schoukens, J. (1999).
\newblock {Measurement and identification of nonlinear systems consisting of
  linear dynamic blocks and one static nonlinearity}.
\newblock \emph{IEEE Transactions on Automatic Control}, 44(6), 1266--1271.

\bibitem[{{Wills} and {Ninness}(2008)}]{Wills2008}
{Wills}, A. and {Ninness}, B. (2008).
\newblock On gradient-based search for multivariable system estimates.
\newblock \emph{IEEE Transactions on Automatic Control}, 53(1), 298--306.

\bibitem[{Zhou et~al.(1996)Zhou, Doyle, and Glover}]{Zhou1996}
Zhou, K., Doyle, J.C., and Glover, K. (1996).
\newblock \emph{Robust and Optimal Control}.
\newblock Prentice-Hall, Inc., Upper Saddle River, NJ, USA.

\end{thebibliography}
               
\end{document}